\documentclass[sigconf,9pt]{acmart}

\usepackage[english]{babel}
\usepackage[ruled,vlined]{algorithm2e}
\usepackage{enumitem}
\setlist{nolistsep}
\usepackage{gensymb}
\graphicspath{{gfx/}}

\newcommand{\fref}[1]{Fig.~\ref{#1}}

\copyrightyear{2020}
\acmYear{2020}
\setcopyright{rightsretained}
\acmConference[EuroP4'20]{3rd P4 Workshop in Europe}{December 1, 2020}{Barcelona, Spain}
\acmBooktitle{3rd P4 Workshop in Europe (EuroP4'20), December 1, 2020, Barcelona, Spain}
\acmDOI{10.1145/3426744.3431321}
\acmISBN{978-1-4503-8181-9/20/12}

\begin{document}

\title{A P4 Data Plane for the Quantum Internet}

\author{Wojciech Kozlowski$^{1,2}$, Fernando Kuipers$^3$, and Stephanie Wehner$^{1,2}$}
\affiliation{%
  \institution{$^1$QuTech, Delft University of Technology, Delft, The Netherlands}
  \institution{$^2$Kavli Institute of Nanoscience, Delft University of Technology, Delft, The Netherlands}
  \institution{$^3$Delft University of Technology, Delft, The Netherlands}
}
\email{{w.kozlowski, f.a.kuipers, s.d.c.wehner}@tudelft.nl}

\renewcommand{\shortauthors}{Wojciech Kozlowski, Fernando Kuipers, and Stephanie Wehner}

\begin{abstract}
  The quantum technology revolution brings with it the promise of a quantum
  internet. A new --- quantum --- network stack will be needed to account for
  the fundamentally new properties of quantum entanglement. The first
  realisations of quantum networks are imminent and research interest in
  quantum network protocols has started growing. In the non-quantum world,
  programmable data planes have broken the pattern of ossification of the
  protocol stack and enabled a new --- software-defined --- network software
  architecture. Similarly, a programmable quantum data plane could pave the way
  for a software-defined quantum network architecture. In this paper, we
  demonstrate how we use P4$_{16}$ to explore abstractions and device
  architectures for quantum networks.
\end{abstract}



\maketitle

\section{Introduction}

The idea of a quantum internet has been around for some
time~\cite{briegel_quantum_1998-1, van_meter_quantum_2014} and in the last few
years physicists have made significant progress towards building the first
long-range quantum networks~\cite{humphreys_deterministic_2018,
  yin_satellite-based_2017, yu_entanglement_2020}. As the hardware grows in
maturity, research interest in software and network architectures for a quantum
internet has also been growing~\cite{dahlberg_link_2019, matsuo_quantum_2019,
  pirker_quantum_2019}. Software-defined networking (SDN) concepts have even
already been applied to quantum key distribution (QKD)
networks~\cite{aguado_enabling_2020}. However, these networks are
single-purpose and are not designed for applications beyond QKD\@. SDNs for
more general quantum networks based on quantum
repeaters~\cite{van_meter_quantum_2014} have not yet been considered. Recent
advances in non-quantum (classical) networking have shown that programmable
data planes offer a powerful foundation for
SDNs~\cite{noauthor_software-defined_nodate}. A programmable quantum data plane
could similarly provide the building blocks for a software-defined quantum
network (SDQN) architecture.

In this paper, we present a software package~\cite{netsquid_qp4}, developed for
NetSquid~\cite{noauthor_netsquid_nodate}, to run P4 programs on a simulated
quantum network. NetSquid is one of the most realistic quantum network
simulators and has already been used to validate new quantum
protocols~\cite{dahlberg_link_2019}. NetSquid's accurate results and rich
library of hardware models mean that P4 programs validated in simulation can
later be ported to real quantum hardware by making the device P4 programmable.
We demonstrate how we use this package to explore quantum data plane
abstractions by implementing a recent quantum
protocol~\cite{dahlberg_link_2019} in P4. Quantum networks are complex and
generally require advanced knowledge of quantum mechanics. With our P4 package,
we start abstracting these low-level details behind quantum device
architectures. In addition to laying the foundation for SDQNs, this could also
make quantum protocol design more accessible.

\section{Domain-Specific Language}

\begin{figure}[t!]
  \centering \includegraphics[width=0.8\linewidth]{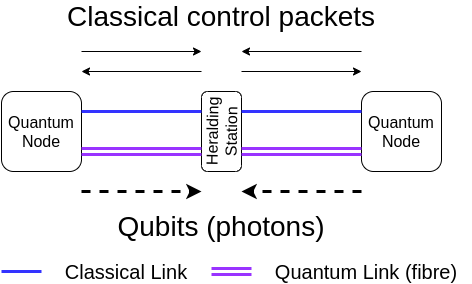}
  \caption{The heralding station upon simultaneous receipt of photons
    ``heralds'' the success or failure of an attempt. }\label{fig:link}
\end{figure}

Classical networks deliver packets from a source to a destination. Quantum
networks distribute quantum entanglement between two or more nodes.
Entanglement is a special state of two or more quantum bits (qubits) in which
the individual qubits cannot be described independently of the others across,
in principle, arbitrary distances. It is the key ingredient for long-distance
quantum communication, because an entangled pair of qubits can be used to
teleport (transmit) arbitrary quantum data.

An entanglement based programmable data plane will most likely need its own
domain-specific language (DSL), just like P4 was created as a DSL for packet
data planes. However, whilst entangled qubits are the fundamental unit of
quantum networks, all protocol control information (i.e.~headers) is
transmitted in classical packets~\cite{dahlberg_link_2019}, which could be
easily processed by a P4 program. Thus, the P4 language offers a convenient
starting point for programmable quantum data planes. The P4$_{16}$ language has
two additional features that make this a practical approach. First, P4$_{16}$
allows the vendor to provide a custom architecture-specific API, which allows
us to define a new architecture that can support a quantum network stack.
Second, P4 has a sizeable ecosystem of open-source software, such as P4Runtime
and ONOS, that will facilitate the deployment of the first experimental SDQNs.

\section{Architecture of a Quantum Network Protocol}

For our demonstration, we have implemented the Midpoint Heralding Protocol
(MHP)~\cite{dahlberg_link_2019}. The architecture of the link is shown in
\fref{fig:link}. The MHP is responsible for the individual attempts to generate
entangled pairs of qubits between two neighbouring nodes. It is a lightweight
protocol that, together with quantum hardware, constitutes the lowest
(i.e.~physical) layer of a quantum network. The next layer in the quantum
protocol stack (the link layer) uses the physical-layer protocol to keep making
attempts until entanglement generation succeeds to provide a robust
entanglement generation service for the end-to-end quantum network layer
service. End-to-end entanglement between two hosts connected to a quantum
network is generated by combining many such link-pairs using a process called
entanglement swapping~\cite{briegel_quantum_1998-1}, but end-to-end
connectivity is beyond the scope of this demonstration.

On each attempt (synchronised between the nodes using a hardware clock and
protocol), the two nodes each emit a photon towards a midpoint station that is
placed in between the two nodes. This midpoint, also called the heralding
station, stochastically succeeds or fails in the entanglement attempt and
``heralds'' the result back to the nodes. If the attempt was successful the two
nodes now each have a qubit that is entangled with its counterpart at the other
node. The MHP will also send a classical message along with the photon towards
the midpoint carrying control information about the photon. The heralding
station responds to this message with a success or failure or an error
notification if the control information from the two nodes was found to be
inconsistent.

\section{Entanglement Generation in P4}

\begin{figure}[t!]
  \centering
  \includegraphics[width=\linewidth]{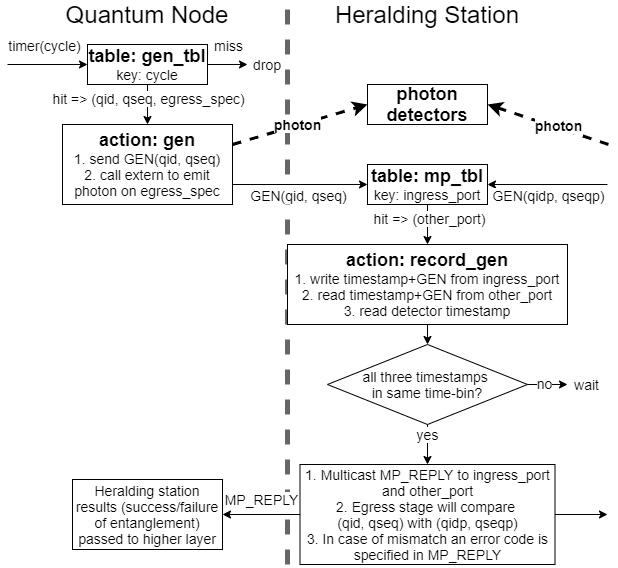}
  \caption{The node upon receipt of the timer pseudo-packet looks up its
    \texttt{gen\_tbl}. A hit yields the parameters for the attempt and the node
    sends a photon and a GEN packet. The heralding station must wait for a GEN
    packet from both sides before it can send an \texttt{MP\_REPLY}. The
    \texttt{mp\_tbl} provides the \texttt{other\_port} value as one heralding
    station can serve more than two nodes. For clarity not all pathways are
    shown.}\label{fig:pipeline}
\end{figure}

To implement the MHP in P4$_{16}$, we wrote two programs: one for the two nodes
on either side of the link and one for the heralding station. We also defined a
new architecture by extending the \emph{v1model} with extern functions for
sending qubits (photons) and multicasting the heralding response from the
midpoint. The P4 pipelines are illustrated in \fref{fig:pipeline}. The MHP
protocol triggers based on a hardware timer that injects a pseudo-packet on
port $0$ at regular intervals. These packets carry a cycle number that is
incremented with every attempt. The P4 program then performs a table lookup to
determine the parameters of the request for this cycle. If the lookup results
in a hit, a photon is emitted on the quantum interface indicated in the request
using one of the new extern functions. Additionally, the program appends the
request information to the timer packet and forwards it via the classical
interface paired with the quantum interface on which the photon was emitted.

The heralding station program is more complicated as it must correlate three
messages. First, the photons will trigger a success/failure notification from
the hardware detectors. Second, the two MHP packets will arrive on either side
with control information that must be processed by the midpoint. The P4 program
achieves this by saving the packet content and the arrival time into P4
registers. Once recorded, it uses the timestamps to check if the other two
messages were received in the same time bin. If so, it multicasts a response to
the two nodes with an indication of success/failure.

\section{Conclusions}

We have presented a software package~\cite{netsquid_qp4} for running P4
programs on a simulated quantum network and we have given an example of an
implementation of a recent quantum network protocol. We plan to use this
framework to explore quantum device architectures and quantum network
abstractions with the intention of developing a software-defined architecture
for quantum networks.

\begin{acks}

We would like to thank Belma Turkovic for helpful discussions. We are also
grateful to Bruno Rijsman for technical discussions and contributions to the
source code of the package.

This work was supported by the \grantsponsor{qt.flagship}{EU Flagship on
  Quantum Technologies}{https://qt.eu/}, \grantsponsor{qia}{Quantum Internet
  Alliance}{https://quantum-internet.team/} (No.~\grantnum{qia}{820445}) (WK,
SW), an \grantsponsor{ercsg}{ERC Starting Grant}{} (SW), and an
\grantsponsor{nwovidi}{NWO VIDI Grant}{} (SW).

\end{acks}

\bibliographystyle{ACM-Reference-Format}
\bibliography{reference}

\end{document}